\documentclass[aip,apl,amsmath,amssymb,reprint,superscriptaddress]{revtex4-1}
\usepackage{graphicx}
\usepackage{amsmath}
\usepackage{dcolumn}
\usepackage{bm}
\usepackage{bbold}

\bibstyle{apsrev4-1}

\begin{document}


\title{Extremely Low-Loss Acoustic Phonons in a Quartz Bulk Acoustic Wave Resonator at Millikelvin Temperature}

\author{Maxim Goryachev}
\affiliation{ARC Centre of Excellence for Engineered Quantum Systems, University of Western Australia, 35 Stirling Highway, Crawley WA 6009, Australia}
\affiliation{Department of Time and Frequency, FEMTO-ST Institute, ENSMM, 26 Chemin de l'\'{E}pitaphe, 25000, Besan\c{c}on, France}

\author{Daniel L. Creedon}
\affiliation{ARC Centre of Excellence for Engineered Quantum Systems, University of Western Australia, 35 Stirling Highway, Crawley WA 6009, Australia}

\author{Eugene N. Ivanov}
\affiliation{ARC Centre of Excellence for Engineered Quantum Systems, University of Western Australia, 35 Stirling Highway, Crawley WA 6009, Australia}

\author{Serge Galliou}
\affiliation{Department of Time and Frequency, FEMTO-ST Institute, ENSMM, 26 Chemin de l'\'{E}pitaphe, 25000, Besan\c{c}on, France}

\author{Roger Bourquin}
\affiliation{Department of Time and Frequency, FEMTO-ST Institute, ENSMM, 26 Chemin de l'\'{E}pitaphe, 25000, Besan\c{c}on, France}

\author{Michael E. Tobar}
\email{michael.tobar@uwa.edu.au}
\affiliation{ARC Centre of Excellence for Engineered Quantum Systems, University of Western Australia, 35 Stirling Highway, Crawley WA 6009, Australia}

\date{\today}


\begin{abstract}
Low-loss, high frequency acoustic resonators cooled to millikelvin temperatures are a topic of great interest for application to hybrid quantum systems. When cooled to 20 mK, we show that resonant acoustic phonon modes in a Bulk Acoustic Wave (BAW) quartz resonator demonstrate exceptionally low loss (with $Q$-factors of order billions) at frequencies of 15.6 and 65.4 MHz, with a maximum $f.Q$ product of 7.8$\times10^{16}$ Hz. Given this result, we show that the $Q$-factor in such devices near the quantum ground state can be four orders of magnitude better than previously attained. Such resonators possess the low losses crucial for electromagnetic cooling to the phonon ground state, and the possibility of long coherence and interaction times of a few seconds, allowing multiple quantum gate operations.
\end{abstract}

\maketitle

To achieve the operation of hybrid mechanical systems in their quantum ground state, it is vital to develop acoustic resonators with very low losses at temperatures approaching absolute zero. The frequencies accessible using mechanical systems are low, and thus a lower temperature is required (governed by $\hbar\omega > k_B T$) to reach the equilibrium ground state \cite{Schliesser2010}. The average number of thermal phonons should follow the Bose-Einstein distribution 
\begin{equation}
\label{eq:boseeinstein}
n_{\text{TH}} \sim \frac{1}{e^{\frac{\hbar \omega}{k_B T}}-1},
\end{equation}
which dictates, for example, that a frequency of greater than 144 MHz is required to have $n_{\text{TH}} < 1$ at a temperature of 10 mK. Recent years have seen mechanical resonators developed with frequencies as high as 11 GHz \cite{TomesPRL2009}, corresponding to $T=0.53$ K for an average phonon occupation number below unity. Aside from conventional thermodynamic cooling, the ground state or Standard Quantum Limit could potentially be reached by exploiting an extremely low loss resonance, where the mode can be cold damped through feedback or a parametric processes. In the ideal case, the ratio $Q/T$ remains constant, where the acoustic $Q$-factor is reduced as the mode amplitude is then damped and electromagnetically cooled to a lower temperature. Alternatively, if the change in state of the resonance can be measured at a rate faster than the dissipation rate $\gamma$, the bath noise is filtered by the high-$Q$ resonance, which is no longer in equilibrium with the bath. In such a case, the effective noise temperature is reduced so that a change in state of order one acoustic phonon could be measureable \cite{Braginsky1992}. The effective temperature, $T_{\text{eff}}$, is then related to the physical temperature by $T_{\text{eff}} = T \tau \gamma /2$, where $\tau$ is the measurement time. With these techniques in mind, the development of cold, high frequency, ultra-low-loss acoustic resonators represents a crucial step in overcoming the challenge of maintaining long coherence times in systems occupying their quantum ground state.\\

Measurement of an acoustic resonator in or near the ground state has only been recently achieved, with results reported by several laboratories \cite{OConnell:2010fk,Teufel:2011kx,Chan:2011uq,Riviere:2011vn,Rocheleau:2010ys}. Such experiments have been performed in a number of novel ways, with quantum limited readout attained by coupling a superconducting Josephson phase qubit to a Thin Film Bulk Acoustic Wave (FBAR) piezoelectric resonator, sideband cooling of a thin aluminium membrane coupled capacitively to a superconducting microwave cavity, trapping of optical modes in a nanofabricated silicon beam, sideband cooling of a micromechanical acoustic mode in a silica Whispering Gallery mode optical resonator, and parametric cooling of a Nb-Al-SiN beam resonator. Notably, all of these devices have an acoustic Q-factor of order $10^4$ or less near the ground state. Historically, the highest acoustic $Q$-factor resonators have been developed for gravity wave detection applications, and are large devices with very low resonant frequencies. Extremely high $Q$-factors from $10^8$ to greater than $10^9$ are readily achievable in bulk sapphire\cite{systemssmalldissipation,lockeparametric,lockeparametric2}, niobium\cite{tobar1993parametric}, and silicon\cite{siliconQ} close to \hbox{4 K}, which has lead to the suggestion that large masses of greater than kilogram scale and low frequency (of order kHz) could potentially be cooled to their ground state via a parametric readout \cite{Tobar2000520}.\\

Recently, high frequency milligram-scale piezoelectric Bulk Acoustic Wave (BAW) quartz resonators were measured down to 4 K with a $Q$-factor of 3$\times 10^8$ at 15.6 MHz \cite{galliou:091911,Galliou:2008ve}. These resonators are advantageous in many respects, namely that they simultaneously exhibit both high frequency and large $Q$-factor, and have strong electromagnetic coupling through the piezoelectric effect. In general, the electromechanical coupling is made more difficult at higher frequencies due to smaller associated displacements for the sensing transducer. However, as in the work of O'Connell et al. \cite{OConnell:2010fk}, the piezoelectricity allows strong electromechanical coupling between photons and acoustic phonons, making them well suited for hybrid quantum applications. Measurements of the BAW resonators at cryogenic temperatures showed a $T^6$ dependence for the Q-factor of the longitudinal mode, and a $T^4$ law for shear modes in the range of 5-21 K, as predicted by Landau-Rumer theory\cite{landaurumer1,landaurumer2}. However, between 3 and 6 K the exponent of the power law decreased, suggesting an additional unknown source of loss.\\

\begin{figure}[t!]
\centering
\includegraphics[width=3.25in]{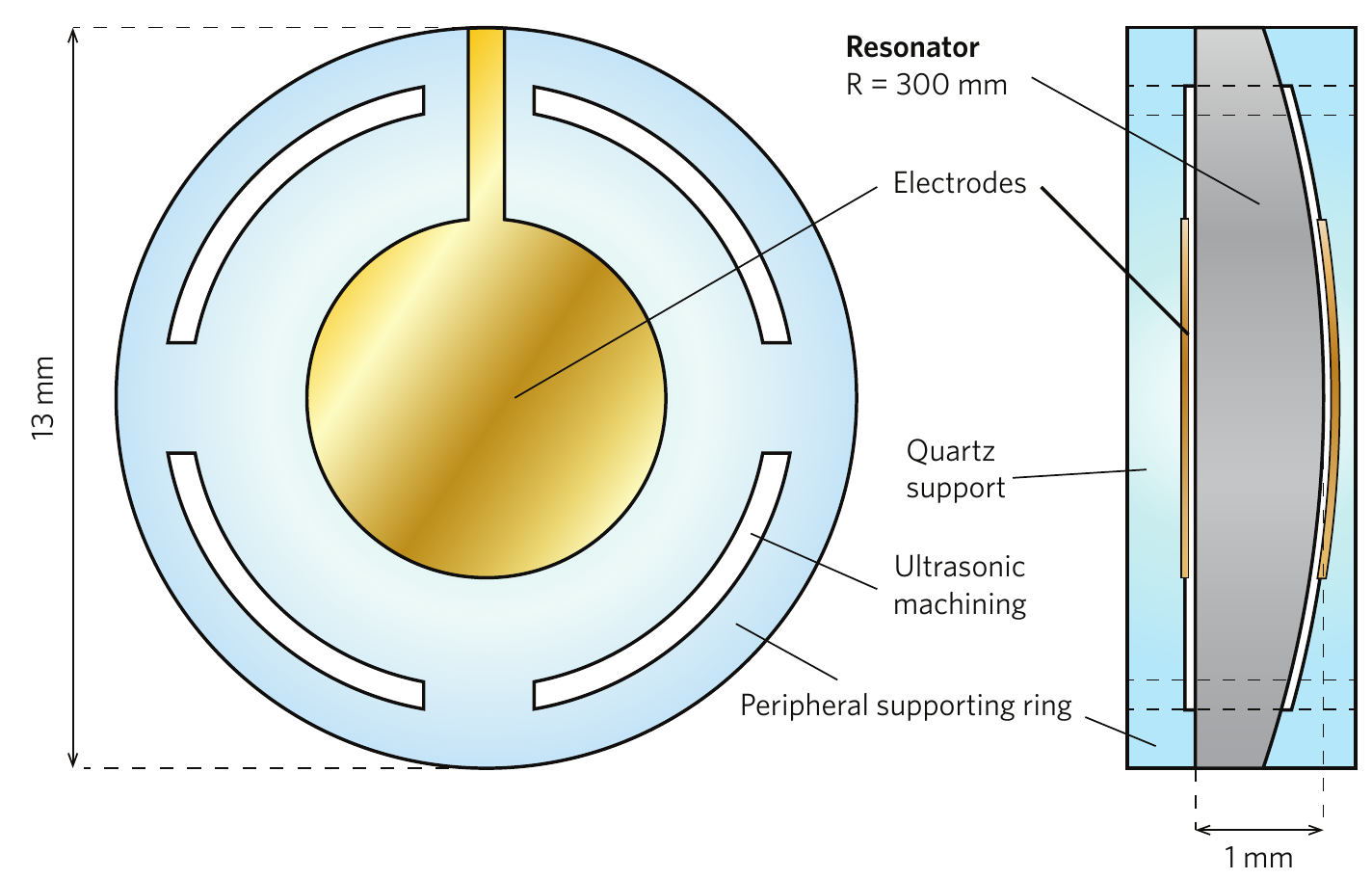}
\caption{\label{resonator}Schematic of the quartz Bulk Acoustic Wave (BAW) resonator manufactured by BVA Industries with non-contacting electrodes. The radius of curvature of the resonator is 300 mm. Vibrations are trapped in the area of the resonator between the electrodes attached to the quartz support structure.}
\end{figure}

In this work, we measure a quartz BAW resonator designed with non-contacting electrodes down to 18 mK, and show that the $Q$-factor continues to increase beyond $10^9$, albeit with a smaller power law exponent. The resonator is a state-of-the-art commercial resonator (SC-cut\cite{1536996}, BVA technology \cite{1537081}) designed for room temperature operation in its shear mode (5 MHz with a frequency-temperature turn-over point at 80 deg. C), a schematic of which is shown in Fig. \ref{resonator}. To allow the acoustic modes to be trapped between the electrodes at the centre of the resonator, isolating the mode from mechanical losses due to coupling to the support ring, the resonator is manufactured with a planoconvex shape \cite{stevens:1811}. The longitudinal overtones are the most strongly trapped and thus exhibit particularly high quality factors. Here, we present measurements characterizing the 5$^{\text{th}}$ and 21$^{\text{st}}$ overtones with effective mode masses of order 5 and 0.7 milligram respectively (total resonator mass of 220 mg) at frequencies 15.6 and 65.4 MHz respectively, which exhibit $Q$-factors exceeding $10^9$ corresponding to decay times of order tens of seconds.\\

\begin{figure}[t!]
\includegraphics[width=3.3in]{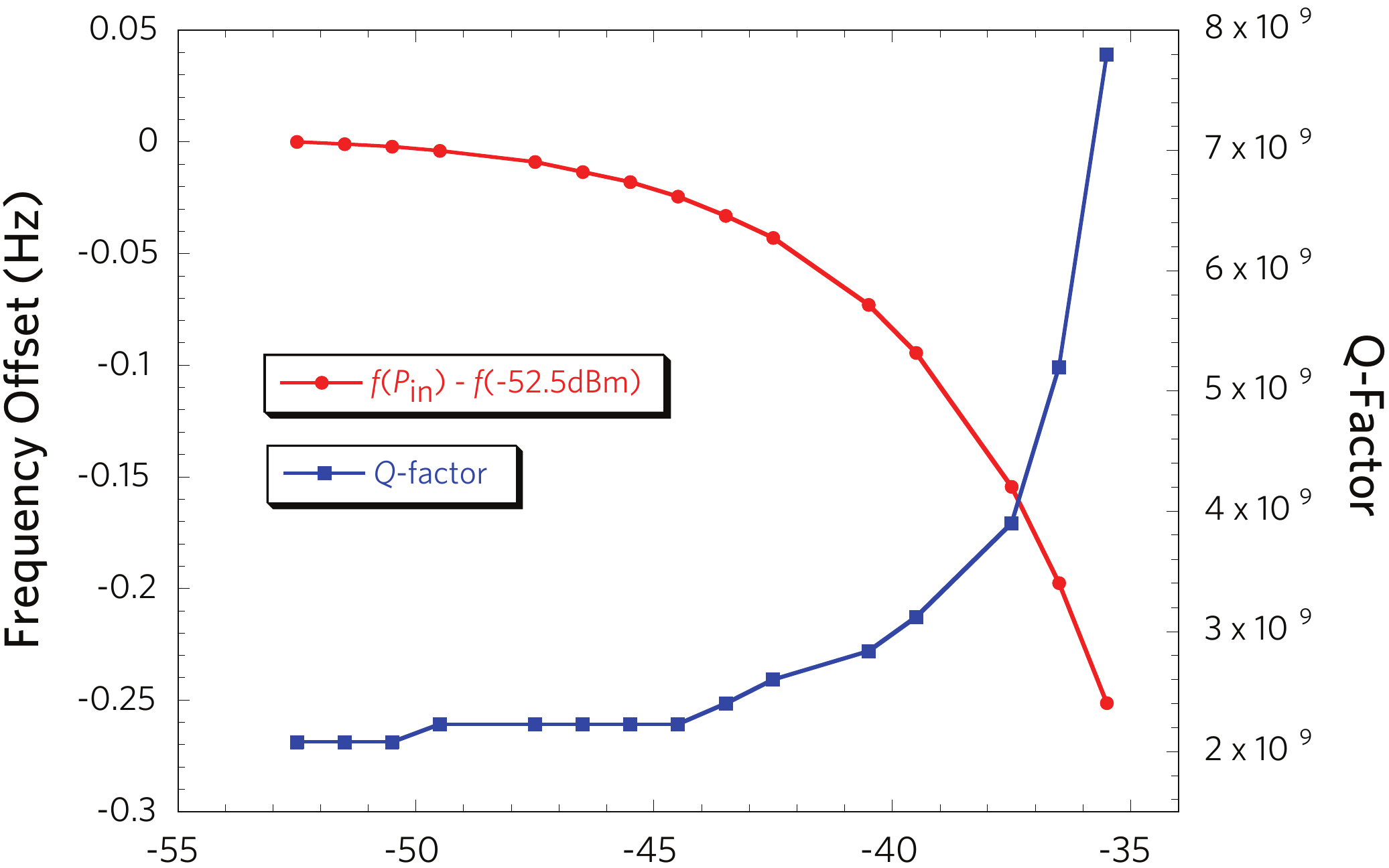}
\caption{\label{powerdependence}Frequency offset versus input power, $f\left(P_{\text{in}}\right)$, at 18 mK for the 5$^{\text{th}}$ overtone with respect to the frequency at -52.5 dBm, $f\left(-52.5\right)=$15,597,316.3845 Hz}
\end{figure}

\begin{table*}[t!]
\caption{\label{table1}Comparison of some acoustic resonator experiments, which have been cooled either to or near the ground state, or have achieved high $Q$-factors. The table shows the frequency, $Q$-factor, environmental temperature, number of thermal phonons ($n_{\text{TH}}$) and $Q$-factor if the system could be cooled ideally to $n_{\text{TH}}=1$ phonon. This assumes that the ratio of $Q/T$  remains constant as the oscillator is simultaneously damped and cooled.}
\begin{tabular}{|l|l|l|l|l|l|}
\hline \textbf{Experiment} & \textbf{Frequency (Hz)} & \textbf{$Q$-factor} & \textbf{Temperature (K)} & \textbf{$n_{\text{TH}}$} & \textbf{$Q (n_{\text{TH}}=1)$} \\ 
\hline Silicon \cite{siliconQ} & 1.96$\times 10^{4}$ & 2$\times 10^{9}$ & 4 & 4.3$\times 10^{6}$ & 677 \\ 
\hline Sapphire \cite{lockeparametric} & 5.33$\times 10^{4}$ & 6$\times 10^{8}$ & 4 & 1.6$\times 10^{6}$ & 554 \\ 
\hline Silica \cite{Riviere:2011vn} & 7.0$\times 10^{7}$ & $10^{4}$ & 0.6 & 179 & 81 \\ 
\hline Spoke-Anchored Silica \cite{Verhagen:2012fk} & 7.8$\times 10^{7}$ & $2.2 \times 10^{4}$ & 0.65 & 1.7 & 83 \\ 
\hline FBAR \cite{OConnell:2010fk} & 6.07$\times 10^{9}$ & 260 & 0.025 & 9$\times 10^{-6}$ & 260 \\ 
\hline Al \cite{Teufel:2011kx} & 1.06$\times 10^{7}$ & 3.3$\times 10^{5}$ & 0.015 & 29.1 & 1.6$\times 10^{4}$ \\ 
\hline Silicon beam \cite{Chan:2011uq,SafaviNaeini:2012ih} & 3.68$\times 10^{9}$ & 4$\times 10^{5}$ & 20 & 113 & 5.1$\times 10^{3}$ \\
\hline Nb-Al-SiN beam \cite{Rocheleau:2010ys} & 6.3$\times 10^{6}$ & $10^{6}$ & 0.10 & 330 & 4.4$\times 10^{3}$ \\ 
\hline Quartz BAW - 5$^{\text{th}}$ overtone & 1.56$\times 10^{7}$ & 2$\times 10^{9}$ & 0.018 & 23.5 & 1.2$\times 10^{8}$ \\ 
\hline Quartz BAW - 21$^{\text{st}}$ overtone & 6.54$\times 10^{7}$ & 1.2$\times 10^{9}$ & 0.021 & 6.2 & 2.6$\times 10^{8}$ \\ 
\hline
\end{tabular} 
\end{table*}

Our experiment was cooled using a cryogen-free Dilution Refrigerator (DR) system, with the quartz resonator thermally connected via an oxygen-free copper mount to the mixing chamber of the DR and cooled to $\sim$20 mK.
To measure the intrinsic $Q$-factor of the quartz BAW resonators, a passive method was employed using an \hbox{Agilent E5061B-005} Impedance Analyzer with a frequency resolution of  0.02 mHz. For BAW devices at low temperatures, this measurement method is preferable to traditional bridge methods\cite{bridge} because of the significant temperature induced changes in the resonator that result in impedance mismatching of the usual $\pi$-network.  A distinctive feature of the impedance analyzer method is a need for compensation of the long connecting cables. For this purpose, three calibration references (an open circuit, short circuit and a 50 $\Omega$ standard) are placed close to the quartz resonator at the end of nominally identical cables. The purpose of the calibration is to remove cable effects from measurement results. Once the calibration is performed under cryogenic conditions for a given frequency range, amplitude, number of points and sweep rate, the magnitude and phase of the resonator motional branch impedance are measured in the vicinity of the resonance. The data is then approximated by a neural network model, and the resonator parameters are obtained through a complex admittance representation of the approximated data.\\

As at 4K \cite{galliou:091911}, the longitudinal mode exhibits superior $Q$-factors to both slow and fast shear modes. For example the 5th overtone of the fast shear mode at 9.2 MHz exhibits a Q-factor of $2.2\times 10^8$, while the 5th overtone of the slow shear mode exhibits a Q-factor of $3.1\times10^8$. Many overtones and anharmonics were measured, with the highest $Q$-factor observed in the 5th overtone longitudinal mode at 15.6 MHz. At high input powers, the resonator undergoes a frequency shift and a significant increase in $Q$-factor, as shown in Fig. \ref{powerdependence} . No distortion of the line shapes was seen in this power range from $ -52.5$ to $-35$ dBm. At higher powers still, close to 0 dBm, nonlinear effects typical in BAW quartz resonators were seen. The $Q$-factor at $-35$ dBm rose to more than 8 billion. This effect is due to a parametric effect that comes from the nonlinearity of a shunt capacitance formed by the resonator electrodes, which are separated by a quartz disk and two gaps. We observe a strong dependence on the excitation level, which has been never been observed at room or liquid helium temperatures in these resonators. Harmonic balance simulations of the corresponding lumped resonator with a nonlinear shunt capacitance model the effect very well, and show an almost exponential increase of the $Q$-factor with a corresponding reduction in frequency.\\
\begin{figure}[b!]
\includegraphics[width=3.3in]{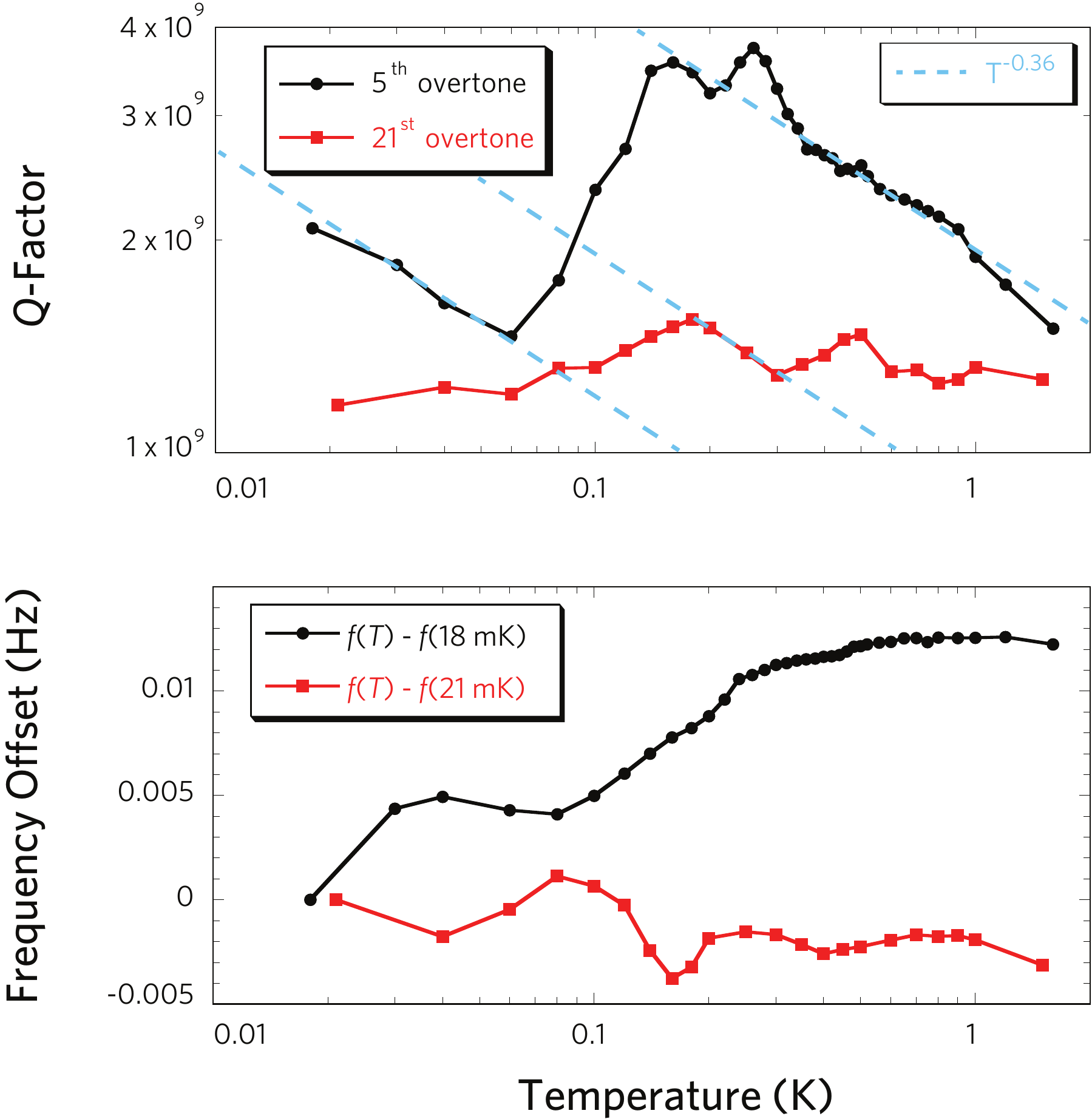}
\caption{\label{qfac}$Q$-factor and frequency offset versus temperature, $f(T)$, for the 5$^{\text{th}}$ and 21$^{\text{st}}$ overtones with respect to the frequency at 18 and 21 mK respectively, at –-52.5 dBm input power. Dashed lines show a $T^{-0.36}$ dependence for both overtones in various temperature regions.}
\end{figure}

At powers below $-50$ dBm, the $Q$-factor remained constant. At this power level, we measured both the 5$^{\text{th}}$ and 21$^{\text{st}}$ overtones as a function of temperature. Figure \ref{qfac} shows the frequency shift and $Q$-factor below \hbox{2 K} in temperature. In this regime of such high $Q$-factor, the loss introduced by the support structure is likely to become important and should be considered. It is also interesting to note that for both modes, the frequency of the overtones exhibit a turning point just below 1 K. At this temperature with such high $Q$-factor and the annulment of the frequency-temperature dependence, a frequency-stable oscillator could be built, which is a point of future investigation. At lower temperatures there are multiple annulment points and no clear dependence of $Q$-factor, although both overtones have peak values between 100 and 300 mK. We also note that in various temperature regions, both overtones exhibit a $T^{-0.36}$ dependence of the $Q$-factor. A power law dependence with the same exponent has been observed previously for the $Q$ of a number of nanoelectromechanical systems such as carbon nanotube resonators \cite{PhysRevLett.93.185501,doi:10.1021/nl900612h}, microresonators in GaAs/InGaP/GaAs heterostructures\cite{shim:133505}, nanomechanical resonators in Si/SiO2/Si\cite{PhysRevB.72.224101}, and micromechanical diamond resonators\cite{PhysRevB.79.125424}.  A theoretical treatment\cite{0295-5075-78-6-60002} calculates a $T^{-1/3}$ dependence due to losses from coupling to so-called ``two level systems", which agrees well with our experimental data. However, the question remains as to why the $Q$-factor degrades in some temperature ranges. Possible explanations could include the coupling of resonance loss mechanisms through the support structure. At such high $Q$-factors, one would expect a greater sensitivity to external loss mechanisms. Comparisons of our experiment with others that have achieved high $Q$-factors at low temperature are summarized in Table \ref{table1}. We compare frequency, $Q$-factor, temperature, thermal phonon occupation number at temperature $T$, and $Q$-factor near the ground state ($n_{\text{TH}}=1$). \\

With such extraordinary $Q$-factors at mK temperatures, quartz BAW resonators are eminently suitable for electromagnetic cooling techniques such as parametric cold damping or feedback cooling. We anticipate that these techniques could be used to prepare the resonator in an ultra-cold quantum state, potentially allowing the resonator to enter the acoustic phonon ground state while maintaining a damped $Q$-factor of more than four orders of magnitude bigger than previously achieved (as suggested in Table \ref{table1}), which corresponds to acoustic decay times of order a few seconds. Also, the resonator exhibits strong electromechanical coupling to mechanical modes through the piezoelectric effect at these temperatures, meaning that a hybrid quantum system is readily achievable. Thus, using BAW technology, single phonon quantum control could be achievable allowing coherence times significantly longer than previously attained, which could allow many more coherent gate operations before coherence is lost with wide ranging implications for quantum information, computing, and control.\\

\begin{acknowledgments}
This work was supported by the Australian Research Council Grant No. CE11E0082 and FL0992016. Collaboration between FEMTO-ST and UWA was also supported by ISL grant number FR100013.
\end{acknowledgments}

\end{document}